\documentclass[%
 preprint, 
 amsmath,amssymb,
 aps, physrev,
]{revtex4-2}

\usepackage{graphicx}% Include figure files
\usepackage{dcolumn}% Align table columns on decimal point
\usepackage{bm}% bold math
\begin{document}

%\preprint{APS/123-QED}

\title{Fluctuation-learning relationship in neural networks}

\author{Tomoki Kurikawa}
 \email{kurikawa@fun.ac.jp}
 \affiliation{Department of Complex and Intelligent Systems, Future University Hakodate, 116-2 Kamedanakano-cho,
Hakodate, Hokkaido, 041-8655, Japan}%Lines break automatically or can be forced with \\

\author{Kunihiko Kaneko}%
\affiliation{
 The Niels Bohr Institute, University of Copenhagen, Blegdamsvej 17, Copenhagen, 2100-DK, Denmark
 }
\affiliation{
 Center for Complex Systems Biology, Universal Biology Institute, University of Tokyo, Komaba, Tokyo 153-8902, Japan
 }
\date{\today}% It is always \today, today,

\begin{abstract}
In neural systems, learning is achieved through a change in synaptic connectivity, following the dynamics in neural activities. 
In general, learning performance depends both on the neural activity and what to be learned.
Experimentally, a possible relation between learning performance and neural activity before learning, in particular, the variability of such neural activity, has been explored.  
Thus, it is essential to establish a general relationship between the learning speed and the variability in order to understand and predict the learning capacity in neural systems.
So far, however, the theoretical basis underlying such a relationship remains to be elucidated. 
Here, following the spirit of the fluctuation-response relation in statistical physics, we theoretically derive two formulae on the relationship between the neural activity and the learning. 
In the first, general formula, the learning speed is proportional to the variance of the spontaneous neural activity and the degree of neural response to the input applied in the learning. 
In the second one that is derived for small input, the speed is proportional to the variances of the spontaneous activities along the target and input directions.
These formulae are shown to hold for various learning tasks with Hebbian or its generalized rules of learning.
We then numerically verify the formulae for input/output mapping and auto-associative memories, which demonstrate that the formulae hold even beyond the regime assumed for the theoretical derivation. 
Although the formulae are derived within a linear regime of the small change in synaptic connectivity, they hold beyond this regime: the formulae estimate the learning time to complete when the synaptic connectivity largely changes.
Our theory also predicts how the learning speed increases with the gain of the activation function of neurons and the number of pre-embedded memories prior to learning, as they increase the variance of the spontaneous fluctuations of the neural activity. 
Furthermore, the formulae also reveal which input/output relationships are feasible to be learned, as is consistent with experimental observations.
Our results, thus,  provide the theoretical basis for the quantitative relationship between the fluctuation of the spontaneous neural activity before learning and the learning speed, which can and will explain a variety of empirical and experimental observations.
\end{abstract}

\maketitle

%\tableofcontents
\section{Introduction}
Learning is a fundamental ability of the brain to adapt to dynamically changing circumstances.
In general, learning performance depends on a large variety of factors in the neural systems\cite{Tang2019,Sheng2022} and the content to be learned\cite{Sadtler2014}.
Learning is achieved by the alteration of neural dynamics that is driven by the change in synaptic connectivity\cite{Abbott2001}.
In turn, the change in connectivity is regulated by the neural dynamics through the activity-dependent plasticity rule (namely, the learning rule).
Thus, the learning results from the interactive changes in both neural dynamics and synaptic connectivity.
To better understand the mechanisms of learning, it is crucial to investigate how the interaction between the changes in neural activities and connectivity leads to learning.

A broad range of recent experimental studies such as motor learning\cite{Wu2014}, generation of bird song\cite{Garrett2013}, and brain-computer interface (BCI)\cite{Hennig2018,Hennig2021,Sadtler2014}, have suggested a postulated relationship between the variability of neural activity before learning and learning speed.
Now it is important to answer a question: how does the variability of the neural activity impact the learning process, and in particular, is there a quantitative relationship between the learning speed and the variability of the neural activity before learning? 

 In statistical physics, it is formulated that the degree of the change in a system's state in response to an external force is proportional to the variance of the state fluctuations in the absence of such a force, as pioneered by Einstein\cite{Einstein1905} and established as fluctuation-response relationship\cite{Onsager1931,Kubo1957,Marconi2008}.
Then, in the spirit of the fluctuation-response relation, one expects that the change in the neural state due to learning, i.e. learning speed, would be correlated with the spontaneous fluctuation of neural activity before learning.

Of course, the neural dynamics are far from thermal equilibrium, and direct application of statistical physics is not available. 
Still, however, in neuroscience, machine learning, and evolutionary biology,  the relationship between fluctuation and response has been proposed following the concept in statistical physics.
In neuroscience,  studies in a rate neuron model and a spiking neuron model \cite{Cessac2019,Deco2023,Nandi2023,Puttkammer2024} demonstrated that the response is represented by the spontaneous fluctuation in neural activities.
Some studies \cite{Yaida2019,Han2021} in machine learning suggested a relation between the learning speed and the fluctuation in the connectivity, while ignoring the variability of neural activity.
In evolutionary biology\cite{Sato2003,Lehner2011,Kaneko2018}, the change rate of phenotype over the generations is shown to be proportional to its fluctuation. 
In contrast, the quantitative relationship between the learning speed and the spontaneous fluctuations of neural activities needs to be elucidated.

In the present study, we first theoretically formulate a general relationship between the learning speed and the spontaneous fluctuations of the neural activity before learning.
According to our relationship, the learning speed is determined by the spontaneous fluctuations in the direction of the target and by the squared neural response to the input.
This relationship is broadly applicable to associative memory in Hebbian learning, I/O mapping in perceptron learning, and other sophisticated machine learning algorithms.

We then verify this relationship numerically using a standard rate neuron model and Hebb-type learning rule for input/output target (I/O) maps and associative memories.
Interestingly, we numerically confirm that this relationship holds much beyond the linear regime
that is originally valid in the initial regime of learning, adopted in the derivation of the formulae.
Hence the validity of the formula we derive is valid over a large range of learning in neural networks in
general.

Our results imply that larger fluctuations in the spontaneous neural activities lead to faster learning, which is consistent with the experimental findings\cite{Sadtler2014}.
The relationship provides a general basis for understanding learning through the interplay between neural dynamics and learning and also paves the way for developing efficient learning strategies.

\section{The fluctuation-learning relationship}
\subsection{The general relationship between spontaneous fluctuations and learning speed}

We investigate the general relationship between spontaneous neural dynamics and speed in learning I/O maps, which includes associative memory as a special case.
A neural network is considered to be trained to generate a target output $\boldsymbol{\xi}$ in the presence of the associated input $\boldsymbol{\eta}$. 
It is composed of $N$ neurons whose activities $\boldsymbol{x}$ evolves as
\begin{align}
    \dot{\boldsymbol{x}} =\phi((J \boldsymbol{x} + \gamma \boldsymbol{\eta})) - \boldsymbol{x} + \boldsymbol{\zeta}, \label{eq:neuro_dyn1}
\end{align}
where $\phi (x)$ is a sigmoid function.
$\boldsymbol{x}$ and $\boldsymbol{\zeta}$  are the neural state and white Gaussian noise, respectively, and they are  $N$-dimensional vectors.
$\boldsymbol{\zeta}$ satisfies  $<\boldsymbol{\zeta} (t_1) \boldsymbol{\zeta}^T (t_2)>=2D\delta(t_1-t_2) I$,
where $<\cdots>$ indicates the temporal average and  $\boldsymbol{\zeta}^T$ is the transpose vector of $\boldsymbol{\zeta}$.
$J$ is a connectivity matrix that represents synaptic connections between neurons, which, we assume, is a full-rank symmetric matrix.
The learning process changes $J$ to make the network memorize an I/O map, i.e., $\boldsymbol{\eta}$ /$\boldsymbol{\xi}$ map.
We assume that the neural dynamics are much faster than the learning, whereas the noise changes much faster than the neural dynamics.
Thus, the neural dynamics are considered to converge into an attractor with a given and constant $J$ and then adiabatically change through the change in $J$.
We consider the regime in which the neural state converges into a fixed point.
The input $\boldsymbol{\eta}$ is applied to the network and the neural state initially converges to an attractor denoted as $\boldsymbol{x_r}$ in the response to $\boldsymbol{\eta}$  with a given initial $J$.
 $\boldsymbol{x_r}$  satisfies
\begin{align}
    \boldsymbol{x_r}=\phi (J \boldsymbol{x_r} + \gamma \boldsymbol{\eta}).\label{eq:response_cond}
\end{align}
Here, the noise term is neglected because $\boldsymbol{x_r}$ is a stationary state wherein the noise effect is temporally averaged out.
In the following analysis of the much slower processes than the noise, such as the fixed points and the learning process, we focused on the dynamics in the absence of the noise effect that is averaged out.
After convergence into $\boldsymbol{x_r}$, the neural state is slowly directed forward to the target state due to the change in the connectivity $\Delta J$ determined by the learning rule.
We analyze this change of the state denoted as $\Delta \boldsymbol{x} = \boldsymbol{x}-\boldsymbol{x_r}$.
By using $J+\Delta J$ instead of $J$ in Eq. \eqref{eq:neuro_dyn1}, 
\begin{align}
    \Delta \dot{ \boldsymbol{x}} = \phi( (J+\Delta J)(\boldsymbol{x_r}+\Delta \boldsymbol{x}) + \gamma \boldsymbol{\eta} ) - (\boldsymbol{x_r}+\Delta \boldsymbol{x}) + \boldsymbol{\zeta}. \label{eq:tmp1}
\end{align}
Now, we consider the initial stage of the learning so that $\Delta J$ and $\Delta \boldsymbol{x}$ are small.
After substituting Eq. \eqref{eq:response_cond} into Eq. \eqref{eq:tmp1} and linearizing it with respect to $\Delta J$ and $\Delta \boldsymbol{x}$  around $\boldsymbol{x}=J \boldsymbol{x_r} + \gamma \boldsymbol{\eta}$ , we obtain the following equation,
\begin{align}
    \Delta \dot{ \boldsymbol{x}} = B (J \Delta \boldsymbol{x} + \Delta J \boldsymbol{x_r}) - \Delta \boldsymbol{x}+\boldsymbol{\zeta}, \label{eq:dyn_deltax}
\end{align}
where $B$ is a diagonal matrix represented by $diag(\beta_1,\beta_2,\cdots,\beta_N)$  with  $\beta_i \triangleq \phi' (J \boldsymbol{x_r} + \gamma \boldsymbol{\eta})_i$.
The fixed point of the dynamics of Eq. \eqref{eq:dyn_deltax}, termed $\Delta \boldsymbol{x}^*$, satisfies
\begin{align*}
    \Delta \boldsymbol{x}^*=  B J \Delta \boldsymbol{x}^* +B \Delta J \boldsymbol{x_r}. 
\end{align*}

Thus, the change in the neural state through the change in the connectivity is formally written as
\begin{align}
    \Delta \boldsymbol{x}^*=(1-B J)^{-1} B\Delta J \boldsymbol{x_r}. \label{eq:lrnspd_deltaJ}
\end{align}

Next, we estimate $(1-B J)^{-1}$ by the neural state, because we cannot know $J$ explicitly in the biological neural system.
If $B$ is sufficiently small such that the spectral radius of $B J$ is less than 1, $(1-B J)^{-1}$ can be estimated with the fluctuation of the neural activity following the equation
\begin{align}
    \dot{\boldsymbol{x}} =-(1-B J) \boldsymbol{x} + \boldsymbol{\zeta}, \label{eq:Langevin}
\end{align}
that corresponds to  the neural dynamics of Eq. \eqref{eq:neuro_dyn1} linearized around $\boldsymbol{x_r}$ in the presence of the input before learning.
Eq. \eqref{eq:Langevin} is the Langevin dynamics with the potential energy   $U(x)=\boldsymbol{x}^T (1-B J) \boldsymbol{x}/2$.
Therefore,  $D (1-B J)^{-1}$ is estimated by the covariance matrix of the stationary distribution of  Eq. \eqref{eq:Langevin}, denoted as $Cov(\boldsymbol{x})_{\text{inp}}$.

By substituting this estimation of $(1-BJ)^{-1}$ into Eq. \eqref{eq:lrnspd_deltaJ}, we obtain  
\begin{align}
    \Delta \boldsymbol{x}^*= D^{-1}Cov(\boldsymbol{x})_{\text{inp}} B\Delta J \boldsymbol{x_r}.  \tag{FLR1} \label{eq:lrnspd_spndelta}
\end{align}
This is a general relationship to relate the learning speed with the covariance of the activity fluctuation before learning.

\subsection{The relationship with Hebb-type learning}
\subsubsection{General form}
Generally, $\Delta J$ is determined by the neural state through a learning rule.
To understand the relationship between the neural state and $\Delta \boldsymbol{x}^*$ further,  we  adopt a Hebb-type learning rule 
\begin{equation}
    \Delta J=(1/\tau_J N)\boldsymbol{f}(\boldsymbol{x})\boldsymbol{g}^T(\boldsymbol{x}) \Delta t \label{eq:learning1}
\end{equation}
as this analysis.
 $\boldsymbol{f}(\boldsymbol{x})$ and  $\boldsymbol{g}(\boldsymbol{x})$ are arbitrary $N$-dimensional functions determining the post- and pre-synaptic contributions in the Hebb learning, respectively.
To make the following analysis clear, we consider $\boldsymbol{g(x)} = \boldsymbol{x}$.
This form widely covers various types of learning rules \cite{Abbott2001}, e.g., those for associative memory\cite{Hopfield1984} and I/O mapping\cite{Kurikawa2013,Kurikawa2020} as noted below.
Note, however,  that even for an arbitrary function of $\boldsymbol{g}(\boldsymbol{x})$, the following analysis holds by replacing $|\boldsymbol{x}|^2$ by $\boldsymbol{g}^T(\boldsymbol{x}) \boldsymbol{x}$.
By substituting Eq. \eqref{eq:learning1} to  \eqref{eq:lrnspd_spndelta},
the initial change $\Delta \boldsymbol{x}^*$, and the initial learning speed $\Delta \boldsymbol{x}^*/\Delta t$, are determined  by the response to the external input and the connectivity as 
\begin{align}
    \Delta \boldsymbol{x}^* / \Delta t=(|\boldsymbol{x_r}|^2/DN\tau_J)  Cov(\boldsymbol{x})_{\text{inp}} B \boldsymbol{f}(\boldsymbol{x}_r). \tag{FLR2-a} \label{eq:lrnspd_res}
\end{align}
This variant of the formula of  \eqref{eq:lrnspd_spndelta} states that the learning speed is proportional to the product of the square of the response and the covariance of the fluctuation before learning.

%Eqs. \eqref{eq:neuro_dyn2} and Eq.\eqref{eq:dyn_deltax} for the change by the external input and by the learning, respectively, are the same forms, namely,  the dynamics of $\boldsymbol{x}$ and $\Delta \boldsymbol{x}$ with the fluctuation are driven by the external force $\beta \gamma \boldsymbol{\eta}$ and $\beta \Delta J \boldsymbol{x_0}$ in addition to the potential energy $U(\boldsymbol{x})$ and $U(\boldsymbol{\Delta x})$, respectively, where $U(x)=x^T (1-\beta J) x^T/2$.
%In the fluctuation-dissipation theorem, the response to the external force is determined by the fluctuation of the spontaneous dynamics.
%Therefore, the responses to the external forces $\boldsymbol{x_0}$ and  $\Delta \boldsymbol{x_0}$ are determined by $\text{diag} (<\boldsymbol{x}^T \boldsymbol{x}>)$ and  $\text{diag} (<\Delta \boldsymbol{x}^T \Delta \boldsymbol{x}>)$.

\subsubsection{Fluctuation-Learning relation for small input}
So far, we considered the general case of the neural dynamics only with the linear regime in the learning, i.e., a small change in $\Delta J$ to obtain  the formula \eqref{eq:lrnspd_res}.
Now, we consider a special case in which the input is small and the response $\boldsymbol{x_r}$ is in a linear regime of $\phi(x)$, 
justified by taking sufficiently small $\gamma$ and a sigmoid function $\phi(x)$ that satisfies $\phi(0)=0$ with its derivative  $\phi$ being small.
In this case, $B=\text{diag}(\beta_1, \beta_2, \cdots, \beta_N) \sim \text{diag}(\beta,\beta,\cdots,\beta)$ with $\beta=\phi'(0)$, so we use the scalar value $\beta$ instead of the matrix $B$ in the following analysis.
Under this assumption, $\boldsymbol{x_r}$ is represented by the covariance of the spontaneous fluctuations.
First, Eq. \eqref{eq:neuro_dyn1} around the fixed point of the origin is approximated as 
\begin{align}
    \dot{\boldsymbol{x}} =-(1-\beta J) \boldsymbol{x} +  \beta \gamma \boldsymbol{\eta}+ \boldsymbol{\zeta}, \label{eq:dyn_lin}
\end{align}
and $\boldsymbol{x_r}$ is the fixed point such as 
\begin{align}
      \boldsymbol{x_r}(\gamma) =  (1-\beta J)^{-1} \beta \gamma  \boldsymbol{\eta}. 
\end{align}
We estimate $(1-\beta J)^{-1}$  by the covariance matrix of the spontaneous fluctuation without the input.
In the following, this covariance is denoted by $Cov(\boldsymbol{x})$ in order to clearly discriminate it from the covariance matrix of the neural fluctuations in the presence of the input $Cov(\boldsymbol{x})_{\text{inp}}$.
Then, we obtain
\begin{align}
    \boldsymbol{x_r} = \frac{\beta \gamma}{D} Cov(\boldsymbol{x}) \boldsymbol{\eta}. \label{eq:resp_spn_eig}
\end{align}
Then, by substituting Eq. \eqref{eq:resp_spn_eig} to  \eqref{eq:lrnspd_res}, we obtain
%\begin{widetext}
\begin{align}
    \frac{\Delta \boldsymbol{x}^*}{\Delta t}  &= \frac{\beta}{DN\tau_J}|\boldsymbol{x_r}|^2  Cov(\boldsymbol{x}) \boldsymbol{f}(\boldsymbol{x}_r)\tag{FLR2-a'} \label{eq:lrnspd_res1} \\
      &= \frac{\beta}{DN\tau_J} (\frac{\beta \gamma}{D})^2 |Cov(\boldsymbol{x})\boldsymbol{\eta}|^2  Cov(\boldsymbol{x}) \boldsymbol{f}(\boldsymbol{x}_r),\tag{FLR3-a} \label{eq:lrnspd_spnspn}
\end{align}
%\end{widetext}
where $Cov(\boldsymbol{x})_{\text{inp}}$ is approximated by  $Cov(\boldsymbol{x})$, because the input is small.
The relation \eqref{eq:lrnspd_spnspn} shows that the learning speed is proportional only to the spontaneous fluctuation without the response.

\subsubsection{Choice of learning rules}
So far, we considered the case in which $\Delta J$ is determined by the Hebb-type rule Eq. \eqref{eq:learning1}.
This rule is a basic and biologically plausible form of synaptic plasticity in the neural system and covers various types of the rule.
Now by specifying  $f(\boldsymbol{x})$ to the following forms, one can rewrite the above fluctuation-learning relationships;
$f(\boldsymbol{x})=\boldsymbol{x}$  corresponding to a simple correlation-based Hebb rule,  
$f(\boldsymbol{x})=\boldsymbol{x}-(\Sigma_i x_i /N)$  to a covariance-based Hebb rule\cite{Abbott2001}, and $f(\boldsymbol{x})=\boldsymbol{\xi}- \boldsymbol{x}$ corresponding to the Perceptron like rule\cite{Kurikawa2013,Kurikawa2016,Kurikawa2020}.

Thus, by using these learning rules, our model applies to different types of memory.
For auto-associative memory, the input pattern $\boldsymbol{\eta}$ is set to be identical to the target pattern $\boldsymbol{\xi}$,
that is, the target itself is applied to the network as input.
Then, the standard model of the auto-associative memory, Hopfield network $J=\Sigma_\mu (\boldsymbol{\xi}^\mu) (\boldsymbol{\xi}^\mu)^T /N$ with its diagonal elements being zero, can be obtained through the correlation-based Hebb rule $f(\boldsymbol{x})=\boldsymbol{x}$ with $\boldsymbol{x}$ clamped by the input $\boldsymbol{\xi}$.

For input $\boldsymbol{\eta}$/output $\boldsymbol{\xi}$ mapping or hetero-associative memory, we adopt the perceptron-like learning $f(\boldsymbol{x})=\boldsymbol{\xi}- \boldsymbol{x}$ or a simple Hebb-like learning $f(\boldsymbol{x})=\boldsymbol{\xi}$.
These learning rules drive $\boldsymbol{x}$ to the target $\boldsymbol{\xi}$ by $\Delta J \boldsymbol{x} = (\boldsymbol{\xi - x})|\boldsymbol{x}|^2$  and $\Delta J \boldsymbol{x} = \boldsymbol{\xi}|\boldsymbol{x}|^2$, respectively.
In fact, the perceptron-like learning was previously demonstrated that this learning rule can memorize sufficiently many I/O maps\cite{Kurikawa2013,Kurikawa2016,Kurikawa2020}.
Consider the initial state of the learning process upon the small input around $\boldsymbol{x_r}$.
Then, $(\boldsymbol{\xi}-\boldsymbol{x}) \sim \boldsymbol{\xi}$ and the perceptron-like learning matches the Hebb-like learning in the initial stage of the learning.

In the following, we focus on learning I/O mappings and train a network to generate a target output $\boldsymbol{\xi}$  in the presence of an input $\boldsymbol{\eta}$.
Thus, we mainly analyze the perceptron-like learning 
\begin{equation}
    dJ/dt = (1/\tau_J N)(\boldsymbol{\xi - x})\boldsymbol{x}, \label{eq:lrn_rule}
\end{equation}
and replace $\boldsymbol{f}(\boldsymbol{x_r})$ by $\boldsymbol{\xi}$ in all variants of the formulae.

The associative memory with the Hopfield network is not the main concern of the paper, 
but even in this case, the formula based on   \eqref{eq:lrnspd_res1} is still valid, as shown in Appendix 4, which indicates that the spontaneous fluctuation along the response pattern is proportional to the learning speed.

\subsubsection{Learning speed for Hebb-type rule determined by the variance of  the spontaneous dynamics}

Now, we rewrite the formula \eqref{eq:lrnspd_res1} by projecting the dynamics onto a basis $X$ that is composed of $N$ unit-vectors.
Then, by  replacing $\boldsymbol{f(x)}$ in the formula \eqref{eq:lrnspd_res1} by $\boldsymbol{\xi}$ and $B$ by $\beta$,  we get
\begin{equation*}
    \frac{\Delta\boldsymbol{x}^*}{\Delta t}=\frac{\beta |\boldsymbol{x_r}|^2}{DN\tau_J} X Cov(X^T \boldsymbol{x}) X^T \boldsymbol{\xi} .\label{eq:lrnspd_P}
\end{equation*}
When the target $\boldsymbol{\xi}$ is parallel to one of the eigenvectors and the basis $X$ is composed of the eigenvectors of $J$, this equation is rewritten\footnote{
    After this projection, the neural dynamics in Eq. \ref{eq:Langevin} are decomposed into $N$ sets of independent 1-dimensional dynamics, where 
    \begin{align*}
    \frac{d (\boldsymbol{p}_i^T \boldsymbol{x})}{dt} =-(1-\beta\lambda_i) (\boldsymbol{p}_i^T \boldsymbol{x}) + (\boldsymbol{p}_i^T\boldsymbol{\zeta}).
\end{align*}
Here, $\boldsymbol{p}_i$ is an eigenvector of $J$ with an eigenvalue $\lambda_i$ and $X=(\boldsymbol{p}_1,,,,\boldsymbol{p}_N )$.
Thus, $XX^T Cov (\boldsymbol{x})XX^T = X \text{diag}( Var_{\boldsymbol{p}_1} (\boldsymbol{x}),,,Var_{\boldsymbol{p}_N} (\boldsymbol{x})) X^T = X \text{diag}((1-\beta \lambda_1)^{-1}, \cdots, (1-\beta \lambda_N)^{-1}  )  X^T$. 
If $\boldsymbol{\xi}=\boldsymbol{p}_i$,  then, $\text{diag}( Var_{\boldsymbol{p}_1} (\boldsymbol{x}),,,Var_{\boldsymbol{p}_N} (\boldsymbol{x}))  X^T \boldsymbol{\xi}= Var_{\boldsymbol{p}_i} (\boldsymbol{x})$.
    } as
\begin{equation}
    \frac{\Delta \boldsymbol{x}^*}{\Delta t}=\frac{\beta |\boldsymbol{x_r}|^2}{DN\tau_J} Var_{\boldsymbol{\xi}} (\boldsymbol{x}) \boldsymbol{\xi}, \tag{FLR2-b} \label{eq:lrnspd_spneig}
\end{equation}
where a scalar value $Var_{\boldsymbol{\xi}} (\boldsymbol{x}) \triangleq (<(\boldsymbol{\xi}^T \boldsymbol{x})^2>-<(\boldsymbol{\xi}^T \boldsymbol{x})>^2)/|\boldsymbol{\xi}|^2$ indicates the variance of the spontaneous fluctuations along the $\boldsymbol{\xi}$ direction.
Even for a random pattern target, by assuming the correlation between the spontaneous activities along different directions is sufficiently small,  $Cov(X^T \boldsymbol{x})$ turns to be a diagonal matrix each element of which is the variance of the spontaneous fluctuation along each line of $X$.
Thus,  the learning speed is approximately represented by the formula \eqref{eq:lrnspd_spneig}.

By applying the same assumption adopted to get Eq. \eqref{eq:resp_spn_eig}, we also obtain a similar representation of  $\boldsymbol{x_r}$,
\begin{eqnarray}
    \boldsymbol{x_r} = \frac{\beta \gamma}{D} Var_{\boldsymbol{\eta}}( \boldsymbol{x}) \boldsymbol{\eta}.
\end{eqnarray}
Then, the formula \eqref{eq:lrnspd_spneig} is transformed into 

\begin{align}
    \frac{\Delta \boldsymbol{x}^*}{\Delta t} = \frac{\beta }{DN\tau_J}
    (\frac{\beta \gamma}{D})^2 (  Var_{\boldsymbol{\eta}} (\boldsymbol{x}) |\boldsymbol{\eta}|)^2 Var_{\boldsymbol{\xi}}(\boldsymbol{x}) \boldsymbol{\xi}. \tag{FLR3-b} \label{eq:lrnspd_spnvar}
\end{align}

Here, we summarize the fluctuation-learning relations we obtained.
First, we derived the general relationship between the learning speed and the spontaneous fluctuations in the formula \eqref{eq:lrnspd_spndelta} that applies to any learning rule.
When we consider the Hebb-type learning rule Eq. \eqref{eq:learning1}, we obtained the formula.\eqref{eq:lrnspd_res}.
Then, assuming the small input and the linear regime of the input,
we obtained the formula \eqref{eq:lrnspd_res1}, which indicates that the learning speed is determined by the covariance matrix of the spontaneous fluctuations and the response, while the formula \eqref{eq:lrnspd_spnspn} does that the speed is determined only by the covariance matrix without the response.
Further, assuming that the spontaneous fluctuations along different directions are uncorrelated, the formulae  \eqref{eq:lrnspd_spneig} and \eqref{eq:lrnspd_spnvar} are obtained.
In  the formula \eqref{eq:lrnspd_spneig}, the learning speed is represented only by the variance of the spontaneous fluctuations along the target and the response in the former relation, while, in the formula \eqref{eq:lrnspd_spnvar}, it is determined by the variance of spontaneous fluctuations along the target and input directions without using the response.
All these relations show how the variance of spontaneous neural fluctuations determines the learning speed.
In the following section, we examine the validity of the (initial) learning speed as $s=|\Delta \boldsymbol{x}^*|/\Delta t$.

\section{Verification of fluctuation-learning relationship with specific connectivity matrices}
To validate the above formulae \eqref{eq:lrnspd_spneig} and \eqref{eq:lrnspd_spnvar}, we consider two specific connectivity models:
a random symmetric Gaussian matrix (random net) model and a pre-embedded-association model\cite{Kurikawa2012}.
We use Eq. \eqref{eq:neuro_dyn1} and Eq. \eqref{eq:lrn_rule} as the neural dynamics and learning rule, respectively.
$\phi(x)$ is set as $\phi (x) = \tanh(\beta x)$.
Only connectivity of the network $J$ is different between these models.
In the following, $N=512$, $2D=10^{-4}$, $\tau_{J} = 100$.
$\gamma=0.001$ for the learning speed,$\gamma=0.1$ for the calculation of the learning time to complete, and  $\gamma=0$ for the spontaneous activity.

\subsection{A random network model}
\subsubsection{Model setting}
Here, we analyze the random network model.
The connectivity $J$ in this model is a random symmetric Gaussian matrix: $<J_{ij}>=0, <(J_{ij})^2>=1/2N, J_{ij}=J_{ji}$.
The eigenvalue distribution $\rho(\lambda)$ of this random matrix follows Wigner's semicircle law\cite{Tao2023} $\rho(\lambda)=\sqrt{2-\lambda^2}/\pi$ when $N \rightarrow \infty$.
Thus, for $\beta < 1/\sqrt{2}$, the origin is a stable fixed point, and the dynamics in Eq \eqref{eq:neuro_dyn1} converge into the origin. We have numerically confirmed the relation between the spontaneous fluctuation and the eigenvalues in Appendix 1.

\begin{figure*}
    \centering
    \includegraphics[width=1\linewidth]{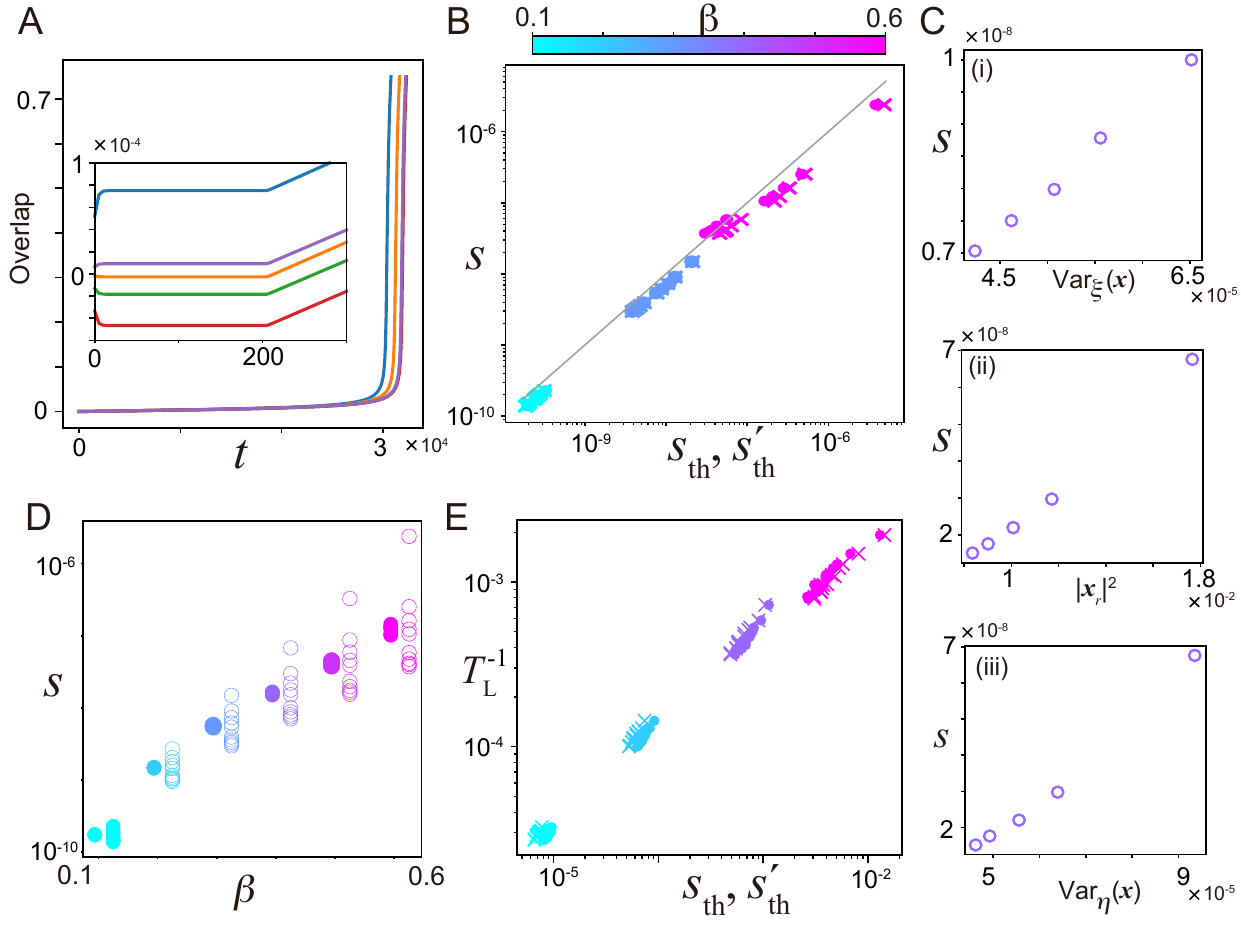}
 \caption{The learning speed for the eigenvector map in the random matrix model.   
    A: The learning process for 5 different maps. The overlaps of the neural state with the targets are plotted.
    B: The measured learning speed $s$ for different values of $\beta$. $s$ is plotted against two types of the theoretical learning speed $s_\text{th}$ of  \eqref{eq:lrnspd_spneig} and $s'_\text{th}$ of \eqref{eq:lrnspd_spnvar} by cross and circle markers, respectively.
    Different color codes the learning speed for different $\beta$ as shown on the color bar. The same color code is in other Panels except for panel A.
    C: $s$ for $\beta=0.4$ is plotted against the variance of the spontaneous fluctuation along the target directions in (i), against the square of response in (ii), and against the spontaneous fluctuation along the input direction in (iii). 
    D: The learning speed as a function of $\beta$ is plotted for the eigenvector map and the random map in the circle outlines and filled circles, respectively.
    E: $T^{-1}_L$ is plotted against $s_\text{th}$ by cross and $s'_\text{th}$ by cross and circle, respectively. The difference in $s_{\text{th}}$ between panels B and D results from the different input strengths. }
    \label{fig:lrnspd_randJ}
\end{figure*}

\subsubsection{Learning speed in the initial stage}
We examine the validity of the fluctuation-learning relation given in the formula \eqref{eq:lrnspd_spneig}. 
To this end, we numerically analyzed the process of learning $\boldsymbol{\xi}$ under the input $\boldsymbol{\eta}$. 
In the learning process, where the connection weight is changed according to Eq.  \eqref{eq:lrn_rule}, the neural activity evolves simultaneously according to Eq. \eqref{eq:neuro_dyn1}.
Figure \ref{fig:lrnspd_randJ}A shows examples of the learning process represented by the overlap of $\boldsymbol{x}$ with the target $\boldsymbol{\xi}$. 
For $t<200$, we run the neural dynamics in the presence of the input without learning to measure the response.
Then, the learning process begins.
The neural state approaches the target monotonically through the change in the connectivity and finally reaches it. 
Here, we measured $|x(t_L+\Delta t)-x(t_L)|/ \Delta t$ as the learning speed $s$ with $t_L=200$ and $\Delta t = 20$.

We numerically examine if the measured learning speed $s$ matches the theoretical  learning speed $s_\text{th}$  determined by the formula \eqref{eq:lrnspd_spneig}, where  $\boldsymbol{x_r}$ is the response defined by $\boldsymbol{x_r}=\boldsymbol{x}(t_L)$.
We have the following two cases of I/O maps:
(i) Eigenvector map: the input and target are eigenvectors of the connectivity matrix.
(ii) Random map: they are random patterns.

For the eigenvector maps, we numerically measured the learning speed for 25 different I/O maps for different values of $\beta$:
We generated two sets of 5 eigenvectors whose eigenvalues are equally spaced in rank and then,  ±- binarized them.
After that, we generated $5 \times 5$ maps by combining these patterns as the inputs and outputs.
The measured learning speed $s$ is plotted in Fig. \ref{fig:lrnspd_randJ}B against the theoretical one $s_\text{th}$.
For all values of $\beta$,  $s$ agrees with the theoretical value $s_\text{th}$ by the formula \eqref{eq:lrnspd_spneig}.
The learning speed is larger and distributed more broadly with the increase in $\beta$.

Now the theoretical value of speed $s_\text{th}$ depends on the variance of the spontaneous fluctuations $Var_{\boldsymbol{\xi}}( \boldsymbol{x})$ and the response $|\boldsymbol{x_r}|$ in  \eqref{eq:lrnspd_spneig}.
To examine explicitly how each of the two factors contributes to the learning speed, we plot the measured speed $s$ as a function of  $Var_{\boldsymbol{\xi}}( \boldsymbol{x})$  and  $|\boldsymbol{x_r}|$ separately, as shown in Fig. \ref{fig:lrnspd_randJ}C.
$Var_{\boldsymbol{\xi}}( \boldsymbol{x})$ is computed for five maps that have identical inputs, whereas  $|\boldsymbol{x_r}|$ is for five maps that have identical targets.
$s$  is linearly proportional to  $Var_{\boldsymbol{\xi}}( \boldsymbol{x})$  and  $|\boldsymbol{x_r}|^2$  well, which agrees with the formula  \eqref{eq:lrnspd_spneig}.

For the random maps, we test the formula  \eqref{eq:lrnspd_spneig} (see  Appendix 2 for the details).
In this case, the learning speed could not exactly match  $s_\text{th}$  due to the approximation of the covariance matrix  by the diagonal variance matrix, and, consequently, the validity of the formula in \eqref{eq:lrnspd_spneig} is not completely assured.
To numerically verify the formula, we computed $s$ and compared it with $s_\text{th}$ in the same manner in case (i).
We found that the measured learning speed $s$ agreed with that predicted by the formula in  \eqref{eq:lrnspd_spneig}.
These findings demonstrate that $s$  is determined by the variance of the spontaneous activity and the response for the general types of I/O maps.

The learning speeds for both types of I/O maps are increased with $\beta$, as shown in Fig. \ref{fig:lrnspd_randJ}D.
The speed for the eigenvector maps has a wider distribution than that for the random pattern case.

We then examine the formula  \eqref{eq:lrnspd_spnvar}, where the initial learning speed is determined only by the spontaneous fluctuation without using the response $\boldsymbol{x_r}$ (denoted as $s'_\text{th}$).
We compare directly $s$ with $s'_\text{th}$ in the same manner as $s_\text{th}$ in Fig. \ref{fig:lrnspd_randJ}B, which confirms the formula  \eqref{eq:lrnspd_spnvar}.
In the same manner as Fig. \ref{fig:lrnspd_randJ}C(i)(ii), we, further, computed $Var_{\boldsymbol{\eta}}( \boldsymbol{x})$ for $5$ maps that have identical targets and found that the learning speed is proportional to $Var_{\boldsymbol{\eta}}( \boldsymbol{x})$ as shown in Fig. \ref{fig:lrnspd_randJ}C(iii).

\subsubsection{Learning time to complete}
So far, we have focused on the initial learning speed that is related to the spontaneous fluctuations.
Can the time to complete the learning of  the target be evaluated by the spontaneous fluctuations?
To answer this question, we numerically calculated the time taken for learning to be completed, $T_L$.
To be specific, $T_L$ is defined as the time when the learning is almost completed; the overlap of the neural state with the target reaches $0.75$, as shown in Fig. \ref{fig:lrnspd_randJ}A.
We examined $T^{-1}_L$  against $s_\text{th}$ in the case of the eigenvector map.

Here, we computed $s_\text{th}$  and found it is proportional to $T^{-1}_L$, as shown in Fig \ref{fig:lrnspd_randJ}E. 
The learning time to complete is successfully estimated by the formula in \eqref{eq:lrnspd_spneig}, although the entire learning process is beyond the small change in the connectivity as assumed to derive this formula.
Further, we found that $T^{-1}_L$ is also proportional to $s'_\text{th}$, indicating that it is evaluated by the formula \eqref{eq:lrnspd_spneig}.
$T^{-1}_L$ in the case of the random map is also proportional to $s_\text{th}$ as well as $s'_\text{th}$, as shown in Appendix 2.

\subsection{Network with pre-embedded I/O map}
\subsubsection{Model setting}
The dynamics in the random network model have no apriori structure before learning a new map, whereas the neural system usually has structured dynamics that were shaped through learning many patterns before learning a new pattern.
The neural system learns a new pattern, depending on the relation between the new and already learned patterns\cite{Duncan2012,Morcos2016}.
To examine if the learning speed formula is valid in this case and how the relation affects the learning process,
we analyze a neural network model in which I/O maps are pre-embedded before a new map is learned.
To be specific, we adopt our previous model\cite{Kurikawa2012} as a canonical sample, which will be explained below.

In this model, the neural dynamics (Eq. \eqref{eq:neuro_dyn1}) and the learning process (Eq. \eqref{eq:lrn_rule}) are the same as the above models, and only the connectivity matrix $J$ is different.
To pre-embedded  input $\boldsymbol{\eta^\mu}$ / target $\boldsymbol{\xi^\mu}$ maps ($\mu=1,2,,,\alpha N$),  $J$ is composed of  $\boldsymbol{\eta^\mu}$ and $\boldsymbol{\xi^\mu}$ as follows\cite{Kurikawa2012}:

\begin{align}
    J &=& (1/N) \Sigma_\mu (\bm{\xi^{\mu}}-\bm{\eta^{\mu}})(\bm{\xi^{\mu}}+\bm{\eta^{\mu}})^T,  \label{eq:J_mem1}
\end{align}
where $\boldsymbol{\eta}^{\mu}$ and $\boldsymbol{\xi}^{\mu}$ are $N$-dimensional random vectors generated in the same manner as the random map in the random network model.
$\alpha$ is the load factor of memories.
It was shown that   $J \bm{\xi^{\nu}} = \bm{\xi^{\nu}} - \bm{\eta^{\nu}}$  and $J \bm{\eta^{\mu}} = \bm{\xi^{\mu}} - \bm{\eta^{\mu}}$ if all patterns of $\{\boldsymbol{\xi}^{\mu} \}$
 and $\{\boldsymbol{\eta}^{\mu} \}$ are mutually orthogonalized;
otherwise $J \bm{\xi^{\nu}} = \bm{\xi^{\nu}} - \bm{\eta^{\nu}} + O(\alpha^{1/2})$  and $J \bm{\eta^{\mu}} = \bm{\xi^{\mu}} - \bm{\eta^{\mu}} + O(\alpha^{1/2})$, where
the last terms of the order of $\alpha^{1/2}$ remains due to the interference between $\{\boldsymbol{\xi}^{\mu} \}$
 and $\{\boldsymbol{\eta}^{\mu} \}$ \footnote{
$(\bm{\xi^{\nu}}-\bm{\eta^{\nu}}) (\bm{\eta}^{\nu})^T \bm{\xi}^{\nu}/N + \Sigma_{\mu \neq \nu} (\bm{\xi^{\mu}}-\bm{\eta^{\mu}})(\bm{\xi^{\mu}}+\bm{\eta^{\mu}})^T \bm{\xi}^{\nu}/N$ for $J\bm{\xi^{\nu}}$ and 
 $(\bm{\xi^{\nu}}-\bm{\eta^{\nu}}) (\bm{\xi}^{\nu})^T \bm{\eta}^{\nu}/N + \Sigma_{\mu \neq \nu} (\bm{\xi^{\mu}}-\bm{\eta^{\mu}})(\bm{\xi^{\mu}}+\bm{\eta^{\mu}})^T \bm{\eta}^{\nu}/N$ for $J \bm{\eta^{\mu}}$ , respectively.}.
Therefore, when $\alpha$ is sufficiently small, the interference terms are negligible so that we obtain $J\boldsymbol{\xi}^{\mu} + \gamma \boldsymbol{\eta^{\mu}} \sim \boldsymbol{\xi^{\mu}} +(\gamma-1)\boldsymbol{\eta^{\mu}} $ and, consequently, the target $\boldsymbol{\xi^{\mu}}$ is a fixed point in the presence of $\boldsymbol{\eta^{\mu}}$ with $\gamma=1$ for $\beta \rightarrow \infty$,  based on the properties of $\tanh(\beta x)$, which holds for all $\mu$.
With the increase in $\alpha$, these interference terms increase, and the fixed points of the targets are unstable.

Our previous studies show that the form of connectivity in Eq. \eqref{eq:J_mem1} is shaped after 
learning $\boldsymbol{\eta}^{\mu}$/$\boldsymbol{\xi}^{\mu}$ maps according to the rule in Eq. \eqref{eq:lrn_rule}\cite{Kurikawa2013,Kurikawa2020}.
Thus, the network model with the connectivity represented in Eq. \eqref{eq:J_mem1}  corresponds to that after learning $\alpha N$ maps, and it is appropriate to study how these pre-embedded maps affect the learning of a new map of $\boldsymbol{\eta}$ and  $\boldsymbol{\xi}$.
In the analysis of this model, the pre-embedded target and input are termed $\boldsymbol{\xi}^{\mu}$ and $\boldsymbol{\eta}^{\mu}$, respectively, while a target and an input to be learned are termed $\boldsymbol{\xi}$ and $\boldsymbol{\eta}$ without supersciript.

Here it should be noted that 
the connectivity matrix $J$ in Eq. \eqref{eq:J_mem1} is neither symmetric nor full-rank and consequently the formulae obtained above do not apply to this model directly.
Thus, we need to examine numerically if the learning speed is evaluated by the formulae \eqref{eq:lrnspd_spneig} and  \eqref{eq:lrnspd_spnvar}.

\begin{figure*}
    \centering
    \includegraphics[width=1\linewidth]{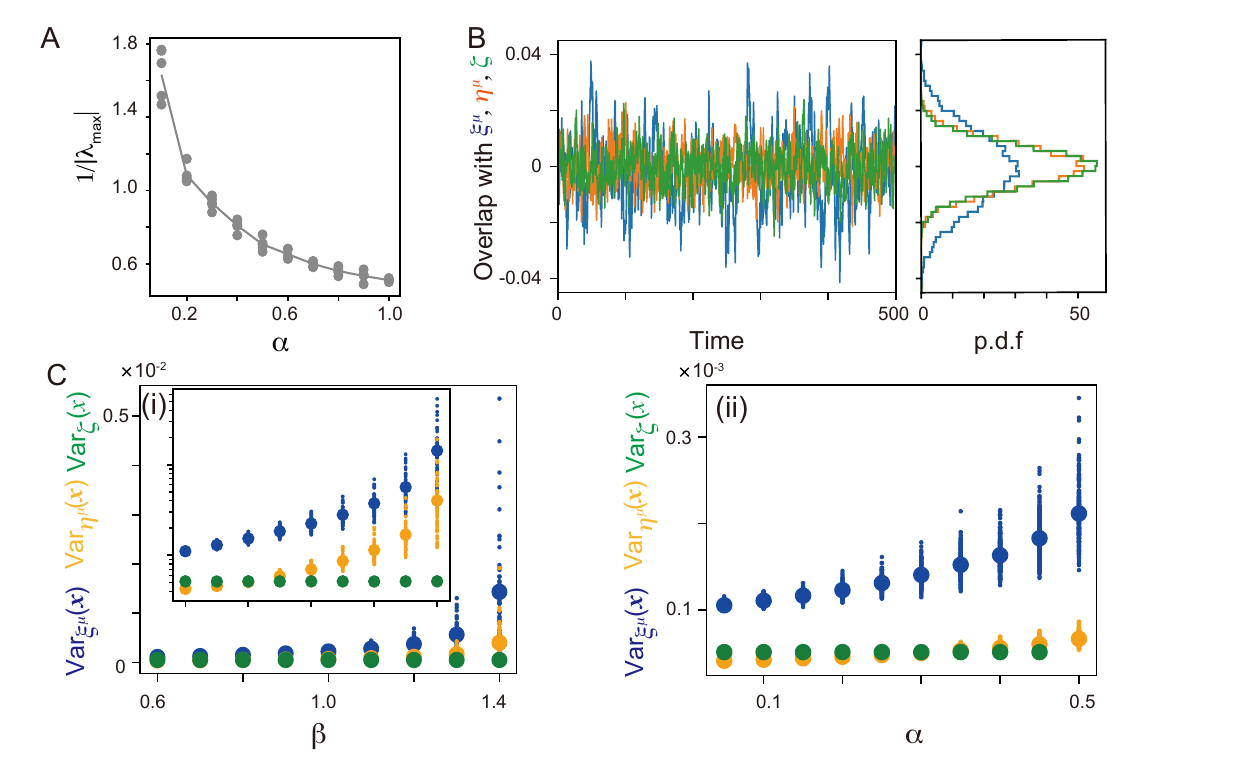}
    \caption{
    A:  $1/|\lambda_{max}|$ is plotted as a function of $\alpha$.  $\lambda_{max}$ indicates the maximum eigenvalue for each realization of the connectivity. These values are calculated for 5 realizations. The full line shows the mean value of $1/|\lambda_{max}|$  across realizations.
    B: Spontaneous fluctuation of the pre-embedded network model in the remapping case for $\alpha=0.36,\beta=0.6$. Left: The spontaneous activity is plotted by using the overlap with a pre-embedded target ($\boldsymbol{\xi}^{\mu}$) and input ($\boldsymbol{\eta}^{\mu}$), and a random pattern ($\boldsymbol{\zeta}$) in blue, orange, and green, respectively. Right: The probability density function of the projected spontaneous activities.
    C: The dependence of the variance of the spontaneous fluctuation with different directions (pre-embedded targets, inputs, and random patterns) on $\beta$ and $\alpha$ in (i) and (ii), respectively. Dots represent the different patterns, while the circle represents the mean value across these different patterns. The inset panel in C(i) is a double logarithmic plot of the original plot of C(i). $\alpha=0.1$ in C(i) and $\beta=0.6$ in c(ii)}
    \label{fig:spn_all}
\end{figure*}

Before analyzing the learning speed, we explore the behavior of the spontaneous fluctuations with $\gamma=0$ by changing $\alpha$ and $\beta$, as they are expected to correlate with the learning speed.
As $\alpha$ increases, the distribution of the eigenvalues is broader, and the variance of the fluctuations also increases.
We focus on the parameter regime $\beta<|\lambda_{max} (\alpha)|^{-1}$, i.e., below the line in Fig. \ref{fig:spn_all}A, in the following analysis, as the trivial fixed point of the origin is stable as postulated in Eq.\eqref{eq:neuro_dyn1}.
$\lambda_{max}(\alpha)$ is the maximum eigenvalue of $J$ when $\alpha N$ maps are pre-embedded.

In Fig. \ref{fig:spn_all}B,  the spontaneous fluctuations are plotted for $\alpha=0.35$ and $\beta=0.6$ by using their projection onto a pre-embedded target $\boldsymbol{\xi}^1$, an input $\boldsymbol{\eta}^1$, and a random pattern $\boldsymbol{\zeta}$ that is orthogonal to all of the pre-embedded patterns.
The fluctuation of the spontaneous activities in the direction of the pre-embedded target is larger than those in the direction of the pre-embedded input pattern and the random one.
Therefore,  the spontaneous fluctuations stretch  more broadly along the target direction than along the other directions.

Such anisotropic property of the spontaneous fluctuations depends on $\alpha$ and $\beta$.
The variances of the spontaneous fluctuations along the pre-embedded targets $\boldsymbol{\xi}^{\mu}$,  $Var_{\boldsymbol{\xi}^{\mu}}( \boldsymbol{x})$  are plotted with the increase in $\beta$ in Fig. \ref{fig:spn_all}C(i), as well as those along the pre-embedded inputs, and their orthogonal patterns, termed   $Var_{\boldsymbol{\eta}^{\mu}}( \boldsymbol{x})$   and  $Var_{\boldsymbol{\zeta}}( \boldsymbol{x})$, respectively.
For any value of $\beta$,  $Var_{\boldsymbol{\xi}^{\mu}}(\boldsymbol{x})$ is larger than the others.
 $Var_{\boldsymbol{\xi}^{\mu}}( \boldsymbol{x})$  and  $Var_{\boldsymbol{\eta}^{\mu}}(\boldsymbol{x})$  rise rapidly as $\beta$ increases, whereas  $Var_{\boldsymbol{\zeta}}( \boldsymbol{x})$   is unchanged.
 Similar behavior is observed for $\alpha$ dependence, as shown in Fig. \ref{fig:spn_all}C(ii).
$Var_{\boldsymbol{\xi}^{\mu}}( \boldsymbol{x})$  and  $Var_{\boldsymbol{\eta}^{\mu}}( \boldsymbol{x})$  increases as $\alpha$ increases, whereas $Var_{\boldsymbol{\zeta}}( \boldsymbol{x})$  is constant.
However, the amount of their change is much smaller than that with increasing $\beta$.

\subsubsection{Initial learning speed and its dependence on $\beta$}
We examine whether the formula \eqref{eq:lrnspd_spneig} is valid in the pre-embedded model.
The learning speed $s$ is computed in the same manner as in the random network model, which is compared with the theoretical learning speed $s_\text{th}$.
We computed the learning speed for $50$ different I/O maps and plot them in Fig. \ref{fig:lrnspd_embJ}A: $25$ maps have identical inputs and different targets and the other $25$ maps have identical targets and different inputs.

Here, the variance of the fluctuations differs depending on the directions of the embedded target and input patterns, and the random patterns.
Therefore, we analyze two cases of I/O maps: (i) remap case: a new map to be learned is set up by combining randomly the pre-embedded $\boldsymbol{\eta^{\mu}}$ and $\boldsymbol{\xi}^{\mu}$ patterns, and (ii) random map case: a new map is chosen from random patterns orthogonal to the pre-embedded patterns.
We investigate the learning speed in these two cases and study its dependence on $\beta$ as well as $\alpha$.

Figure \ref{fig:lrnspd_embJ}A shows $s$ and $s_\text{th}$  for different values of $\beta$ with $\alpha=0.1$ for the remap case.
The values of the learning speed for each $\beta$ are distributed along the diagonal line.
Thus, the formula  \eqref{eq:lrnspd_spneig} is valid,
although their distribution for $\beta=1.4$ deviates from the diagonal slightly more than those for smaller $\beta$\footnote{$\beta=1.4$ is close to the boundary under which the fixed point of the origin is stable and the assumption for the learning speed formula is almost broken.}.
Further, we analyze the contributions of the spontaneous fluctuation and $|\boldsymbol{x_r}|^2$ to the learning speed, separately.
In Fig. \ref{fig:lrnspd_embJ}B, $s$ is shown as a function of  $Var_{\boldsymbol{\xi}}( \boldsymbol{x})$ and $|\boldsymbol{x_r}|^2$, which demonstrates that it is linearly proportional to $Var_{\boldsymbol{\xi}}( \boldsymbol{x})$  and $|\boldsymbol{x_r}|^2$.
 $Var_{\boldsymbol{\xi}}( \boldsymbol{x})$ was computed for $25$ maps that have identical inputs, whereas  $|\boldsymbol{x_r}|^2$ is for $25$ maps that have identical targets.

We then examined the case of the random map (see the detail in Appendix 3), and confirmed that the measured learning speed $s$ agrees again with the theoretical learning speed $s_{\text{th}}$ in the formula \eqref{eq:lrnspd_spneig} for various values of $\beta$.
Thus, for different types of I/O maps, the formula between the spontaneous fluctuation and the learning holds.
We also compare the learning speeds $s$ quantitatively between the two types of I/O maps in Fig. \ref{fig:lrnspd_embJ}C.
The learning speed in both types increases with $\beta$.
For a broad range of  $\beta$, the learning speed in the remap case is higher than that in the random map, reflected in the larger variance of the fluctuation in that case.

\subsubsection{Learning time to complete}
We, next, examine whether the learning time to complete, $T_L$, is estimated by the spontaneous activity and the response like the random network model.
We numerically measured  $T^{-1}_L$ and plotted it against $s_\text{th}$ in Fig. \ref{fig:lrnspd_embJ}D for the remap case and found that $T^{-1}_L$ is clearly proportional to $s_\text{th}$.
For the random map,  $T^{-1}_L$ is also proportional to $s_\text{th}$ (see Appendix 3).
Thus, $T_L$ is also estimated well by the spontaneous fluctuation in the direction of the target to be learned and the response in the pre-embedded network model.

\subsubsection{Dependence of the learning speed on the number of memories $\alpha$}
\begin{figure*}
    \centering
    \includegraphics[width=0.75\linewidth]{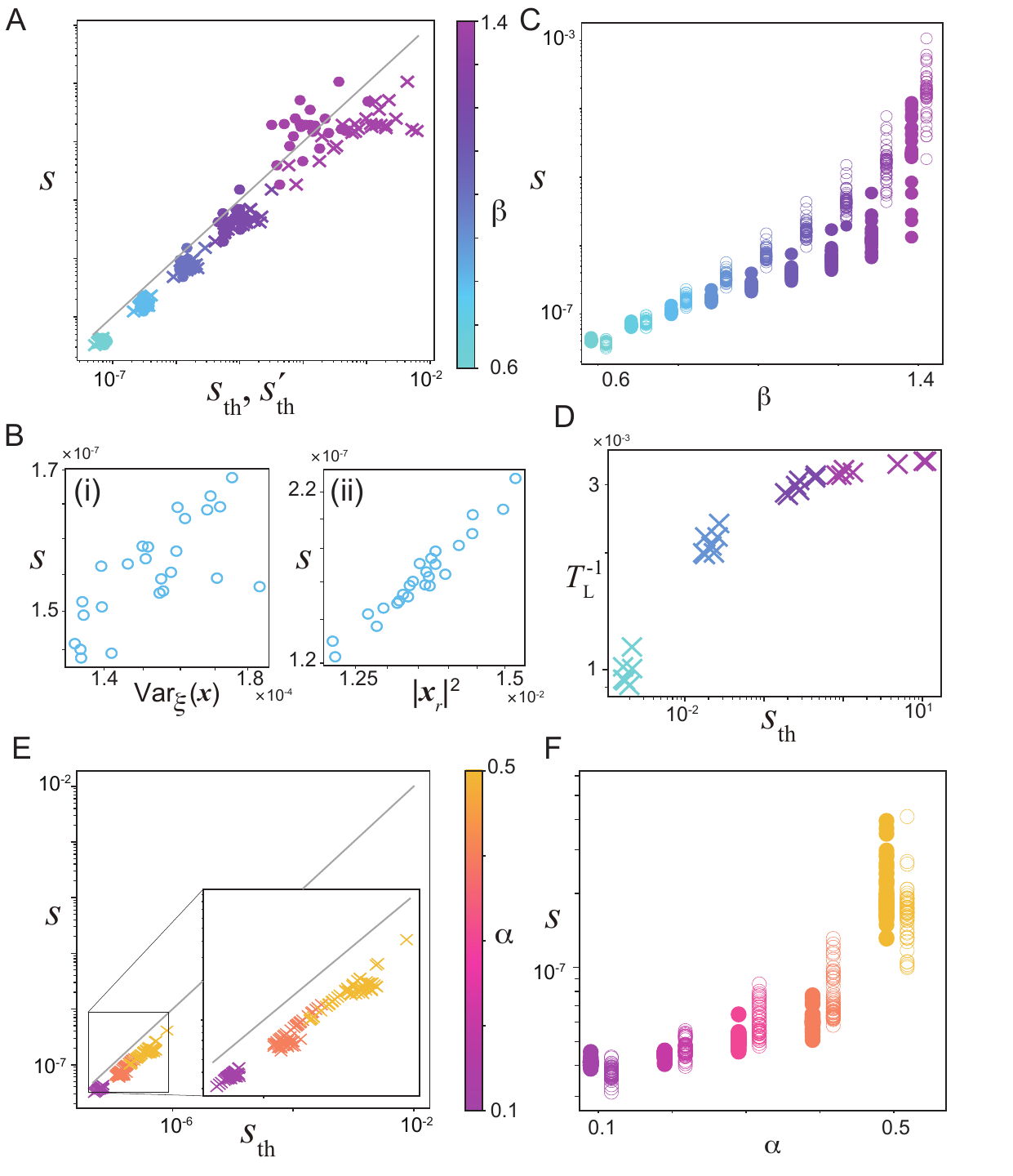}
    \caption{The learning speed in the pre-embedded model in the remap case. 
     A: The learning speed $s$ against the theoretical ones $s_\text{th}$ and $s'_\text{th}$ is depicted in cross and circle respectively, for different values of $\beta$ for $\alpha=0.1$. The color code is shown in the right bar. B: $s$ for different 25 maps and for $\beta=0.6$ are shown against the variance of the spontaneous fluctuations along the target direction in (i) and against the response in (ii).  
 C: The learning speed is plotted as a function of $\beta$ for the random map in full circles and for the remap in outlined circles.
 D: The inverse of the learning time to complete $T^{-1}_L$ is plotted against $s_\text{th}$.
 E: $s$ against $s_\text{th}$ is plotted for different values of $\alpha$ for $\beta=0.6$. The color code is shown in the right bar.
 F: The learning speed is plotted as a function of $\alpha$ in the same manner as in C. }
    \label{fig:lrnspd_embJ}
\end{figure*}

Finally, the dependence of the learning speed on $\alpha$ is investigated.
Figure \ref{fig:lrnspd_embJ}E shows that the learning speed $s$ in the remap case agrees with the theoretical value $s_\text{th}$ predicted by the formula \eqref{eq:lrnspd_res1} across the various values of $\alpha$.
Compared to the distribution of $s$ across various values of $\beta$, that of $s$ across $\alpha$ is narrower.
Also, we found that the learning speed for the random map case is consistent with the formula \eqref{eq:lrnspd_res1}, as shown in Appendix 3.
These results demonstrate that the learning speed formula is valid for different $\alpha$.
Further, figure \ref{fig:lrnspd_embJ}F shows the increase in the learning speed with $\alpha$, reflected by the increase in the spontaneous fluctuations with $\alpha$,  as shown in Fig. \ref{fig:spn_all}C(ii).

\section{Discussion}
We have elucidated the relationship between the spontaneous fluctuation and learning speed in the I/O mapping or the associative memories.
By following the spirit of the fluctuation-response relationship\cite{Einstein1905,Onsager1931,Kubo1957}, the general formulae for 'fluctuation-learning relationship' have been derived, which states that the initial learning speed is determined by the (co-)variance of the spontaneous fluctuations as delineated in the formulae \eqref{eq:lrnspd_spndelta}, \eqref{eq:lrnspd_spneig} and Eq. \eqref{eq:lrnspd_spnvar}.
The most general form of this relationship, as represented in the formula \eqref{eq:lrnspd_spndelta}, indicates that the change in the neural state during learning is determined by the covariance matrix of the spontaneous fluctuation and the connectivity change $\Delta J$ which is valid in any learning method.
Assuming a widely-used Hebb-type learning rule\cite{Abbott2001} in Eq. \eqref{eq:learning1}, a more specific representation of this relationship is obtained as the formula \eqref{eq:lrnspd_spneig}, which shows that the learning speed is determined by the variance of the spontaneous dynamics along the target direction as well as by the input response.
Moreover, by restricting the linear regime around the origin in Eq. \eqref{eq:tmp1}, we have shown that the rate of learning is determined solely by the variance in the spontaneous dynamics along the target and input directions as represented in Eq. \eqref{eq:lrnspd_spnvar}.

We also numerically verified these formulae \eqref{eq:lrnspd_spneig} and \eqref{eq:lrnspd_spnvar} in the random network and pre-embedded models with the perceptron-like rule for learning the I/O mapping, as well as the Hopfield network model for associative memory.
Notably, even in the pre-embedded model — which is characterized by neither asymmetry nor full-rank matrix, thus beyond the assumptions of our formulae  \eqref{eq:lrnspd_spneig} and \eqref{eq:lrnspd_spnvar} — the observed learning speed agrees well with that predicted by these formulae.
For associative memory, the Hopfield network model with Hebb rule was numerically examined in Appendix 4, validating the fluctuation-learning relationship.
Finally, our numerical findings demonstrated that the learning time to complete beyond the initial stage of the learning is also inversely proportional to spontaneous fluctuations, consistent with the expectation by our formula.

Note that in the derivation of our formula, linearization approximations to inputs and noise are adopted.  However, the numerical results demonstrate that the formula is valid in a rather broad range, not restricted to the initial regime.  Such expansion of linear regime is often observed for a stable system.  It is recently noted in the study of evolutionary-fluctuation response both in simulations and experiments\cite{Sato2003,Kaneko2018,Furusawa2018,Maeda2022,Kaneko2024}, whereas the validity of linear-response theory to a macroscopic scale has been noted in statistical physics. Considering the stability of the neural state that keeps memories, one may expect the extension of a linear regime, while its theoretical clarification will be an important future issue.

The uncovered relationship between learning speed and variance in the spontaneous neural activities along the target and input directions has important implications to address the question in which case the learning is facilitated.
First, if the neural dynamics have a larger magnitude of spontaneous activities, such neural dynamics can afford faster learning.
Second, this relationship provides which maps of input and target are more feasible to be learned.
If they are correlated more with spontaneous neural activities, they can be learned more easily. 
Now we discuss these two implications.

First, recent experimental studies have suggested that the larger variability before learning significantly enhances the speed of learning\cite{Wu2014,Tumer2007}.
The motor variability in the arm-reaching task\cite{Wu2014,Dhawale2017} and vocal variability in birdsong learning\cite{Tumer2007} were observed to promote learning, suggesting that variability of neural correlates of the behaviors in the motor cortex\cite{Wei2019,Kondapavulur2022} and song-specialized basal ganglia circuits\cite{Picardo2016} could accelerate learning speed.

In our study, the learning speed increases with the variance of spontaneous fluctuations as the gain $\beta$ or the number of memories $\alpha N$ increases.
Previous theoretical studies demonstrated that the gain parameter is related to the performance in several learning methods\cite{Orhan2019,Kurikawa2020,Kurikawa2021d} and the recall performance\cite{Kurikawa2023}.
Experimental observations show that attention increases the sensitivity of the neural response.
Attention modulates neural dynamics towards the asynchronous state\cite{Harris2011} or the high active state\cite{Buehlmann2008} that are sensitive to external response\cite{Luczak2013}.
As for the relation between the number of memories and learning ability, 
studies in cognitive science demonstrated that having a larger vocabulary enhances the ability to learn new words\cite{Werker2002}, 
yet the underlying neural mechanisms have not been fully elucidated.
Our fluctuation-learning relationship sheds a light on how various factors such as attention, pre-learned vocabulary, and behavioral variability are related to the learning speed in a common framework.

Second, experimental studies\cite{Sadtler2014, Hennig2018,Hennig2021} by using brain-computer interfaces (BCI) directly demonstrated that the learning speed in cursor-movement tasks differed depending on the correlation between the target patterns to be learned and the spontaneous neural dynamics, wherein given neural activity patterns in an animal were directly linked to cursor movement on a computer monitor without the use of its arms or eye movement.
The animal quickly mastered new mappings between cursor movement and a specific neural activity pattern when this pattern was aligned with the direction of higher variance in spontaneous dynamics (on-manifold).
Conversely, when the mapping to be learned was less correlated with the spontaneous neural dynamics (off-manifold), the learning required more time.
Even in the same task, the learning speed depended on the variance of the spontaneous dynamics for different mappings, which agrees with our formula.
A study in machine learning with a recurrent neural network model\cite{Feulner2021} also showed the difference in the learning speed between on- and off-manifold, whereas its mathematical foundation for these differences had yet to be explored.
Recent studies, including these BCI studies\cite{Sadtler2014,Hennig2021},  have suggested a possible relation between the geometry of the spontaneous dynamics and the learning performance\cite{Kurikawa2018}.
Our simple I/O mapping and associative memory models provide the theoretical foundation of such a relation.

The formula  \eqref{eq:lrnspd_spndelta} presents the relation between the learning speed and the change in the connectivity $\Delta J$ under the assumption that the neural dynamics converge into the fixed point wherein the timescales in the neural and learning (synaptic) dynamics are sufficiently separated.
This assumption is generally satisfied, in neural systems, irrespective of learning rules.
Our relationship can apply to various learning methods, such as backpropagation\cite{Rumelhart1986} in machine learning, which computes $\Delta J$ from the derivative of an objective function concerning connectivity, and the reward optimization method in reinforcement learning that determines $\Delta J$ based on the derivative of the reward in relation to connectivity.
Typically, these methods have overlooked the role of dynamic fluctuations in neural activities that modulate the learning rate.
Our study suggests a possible scheme to enhance learning efficiency in these domains.

The fluctuation-response relationship\cite{Einstein1905,Onsager1931,Kubo1957,Marconi2008} generally formulates the relationship between spontaneous fluctuations and system responses to external forces.
The possible relevance of this relation to neuroscience\cite{Cessac2019,Deco2023,Nandi2023,Puttkammer2024} and machine learning\cite{Yaida2019,Han2021} has been studied.
To our knowledge, however, ours is the first to analyze the relationship between spontaneous neural fluctuations and learning speed.
Previous theoretical studies, based on the standard rate-coding neural network\cite{Cessac2019,Nandi2023} and the integrate-and-fire neuron model\cite{Cessac2019,Puttkammer2024}, have suggested a relationship between spontaneous dynamic fluctuations and neural response, whereas the deviation of the neural dynamics from the equilibrium state is observed by using the fluctuation-dissipation theory\cite{Deco2023}.
These studies, however, did not elucidate the relationship between spontaneous fluctuations and learning speed.
Conversely, some studies\cite{Yaida2019,Han2021} in machine learning focused on the link between the learning process and the fluctuations of the connectivity dynamics, but not of neural activities.
Our research contrasts these approaches by examining how spontaneous neural fluctuations impact learning speed.

In evolutionary biology, the extension of the fluctuation-response relationship has been advanced both in theory\cite{Sato2003,Kaneko2018} and experiment\cite{Kaneko2018,Maeda2022}, where the evolution speed, that is, the change rate of phenotype over the generations, is shown to be proportional to its fluctuation.
In the evolution process, the gene regulation network governing the phenotype is modified depending on the phenotype through natural selection over the generations, which is analogous to the neural-activity-dependent learning process.
Then, it will be interesting to pursue the possible link between the fluctuation-learning relationship and an evolutionary fluctuation-response relationship in future studies.

Reservoir computation\cite{Maass2002,Jaeger2004} has been focused on as an efficient learning framework. Some studies demonstrated that specific connectivity structures of the reservoir, such as a log-normal distribution of synaptic weights\cite{Teramae2012,Matsumoto2023}, improved memory capacity. Other studies showed that the edge of chaos\cite{Legenstein2007,Kurikawa2023} or chaotic itinerancy\cite{Kaneko2003,Inoue2020} optimized computational performance.  However, the relationship between spontaneous fluctuations and what is to be learned is almost ignored. Our study suggests a novel approach to promote learning speed; if I/O maps are assigned to the direction of the larger spontaneous fluctuation in the reservoir, these maps could be learned faster.

To sum up, we elucidated the relationship between spontaneous fluctuations and learning speed, which generally holds across several theoretical neural network models with Hebb-type learning rule. The uncovered relationship, thus, provides a general viewpoint on how learning is accelerated by taking advantage of the variance in spontaneous neural activities. 

\section*{Acknowledgments}
We thank Tatsuya Haga for the valuable comments.
The present work is supported by JSPS KAKENHI (No. 20H00123, T.K. and K.K.), NNF21OC0065542 (K.K.), and Special Research Expenses in Future University Hakodate
(T.K.).
K.K. is also supported by MIC under a grant entitled
“R and D of ICT Priority Technology 421 (JPMI00316)”.
%%%%%%%%%%%%%%%%%%%%%%% References %%%%%%%%%%%%%%%%%%%%%%%%%

%Add references with BibTeX or manually 
%\cite{Zhang:14,OPTICA,FORSTER2007,Dean2006,testthesis,Yelin:03,Masajada:13,codeexample}.

%%%%%%%%%% If using BibTeX:
%\bibliography{sample}
\bibliography{16th_refs}
%%%%%%%%%% If preparing manually:
% \newcommand{\enquote}[1]{``#1''}

% \bibitem{Zhang:14}
% Y.~Zhang, S.~Qiao, L.~Sun, Q.~W. Shi, W.~Huang, L.~Li, and Z.~Yang,
%   \enquote{Photoinduced active terahertz metamaterials with nanostructured
%   vanadium dioxide film deposited by sol-gel method,}
%   {\protect\JournalTitle{Optics Express}} \textbf{22}, 11070--11078 (2014).

% \bibitem{Optica}
% {Optica}, \enquote{{Optica Publishing Group},}
%   \url{http://www.opg.optica.org}.

% \bibitem{FORSTER2007}
% P.~Forster, V.~Ramaswamy, P.~Artaxo, T.~Bernsten, R.~Betts, D.~Fahey,
%   J.~Haywood, J.~Lean, D.~Lowe, G.~Myhre, J.~Nganga, R.~Prinn, G.~Raga,
%   M.~Schulz, and R.~V. Dorland, \enquote{Changes in atmospheric consituents and
%   in radiative forcing,} in \enquote{Climate Change 2007: The Physical Science
%   Basis. Contribution of Working Group 1 to the Fourth Assesment Report of
%   Intergovernmental Panel on Climate Change,}  S.~Solomon, D.~Qin, M.~Manning,
%   Z.~Chen, M.~Marquis, K.~B. Averyt, M.~Tignor, and H.~L. Miler, eds.
%   (Cambridge University Press, 2007).

% \end{thebibliography}

\section*{Appendix 1: The spontaneous fluctuation and eigenvectors in the random network model}
\begin{figure}[h]
    \centering
    \includegraphics[width=0.75\linewidth]{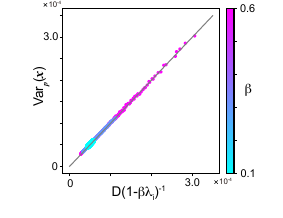}
    \caption{The variance of the spontaneous fluctuation along eigenvectors $p_i$ against $D(1-\beta \lambda_i)^{-1}$. Values for different $\beta$ are shown in different colors.
    The solid line represents the diagonal line consistent with $Var_{\boldsymbol{p_i}(\boldsymbol{x})} = D(1-\beta \lambda_i)^{-1}$.}
    \label{fig:spn_rndJ}
\end{figure}

For the eigenvector maps in the random network model, $(1-\beta J)^{-1}$ in Eq. \eqref{eq:dyn_lin} is transformed to $(1-\beta \text{diag}(\lambda_1,\lambda_2,,,\lambda_N))^{-1}$ by projecting the dynamics onto the basis $X$ composing of the eigenvectors of $J$. 
the eigenvectors and their eigenvalues are denoted as $\boldsymbol{p_i}$ and $\lambda_i$, respectively.
Here, we examine the relation between the spontaneous fluctuation and the eigenvectors represented by $Var_{\boldsymbol{p_i}(\boldsymbol{x})} = D(1-\beta \lambda_i)^{-1}$, as shown in Fig. \ref{fig:spn_rndJ}A.
For the larger $\beta$, the variance of the spontaneous fluctuation along the eigenvector $\boldsymbol{p}_i$ is larger, in proportion to  $(1-\beta\lambda)^{-1}$.
Despite the dependence of the variance of the distribution on $\beta$,
$Var_{\boldsymbol{p_i}}(\boldsymbol{x})$ almost agrees with  $D(1-\beta\lambda_i)^{-1}$ for all of $i$ and for all values of $\beta$.

\section*{Appendix 2: the random patterns in the random network model}

\begin{figure*}
    \centering
    \includegraphics[width=0.7\linewidth]{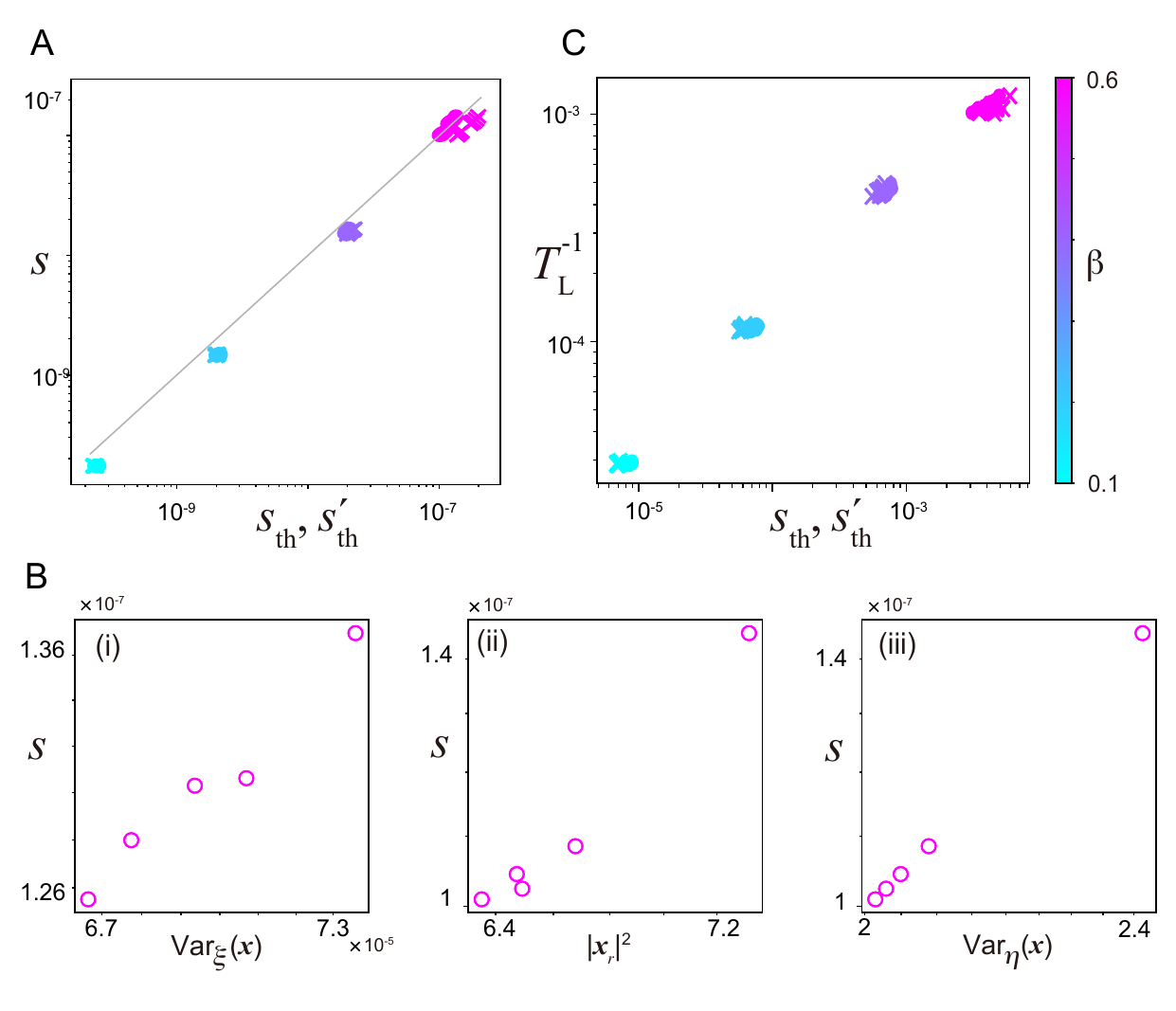}
    \caption{
    The learning speed in the random map case in the random network model.
    A: The measured learning speed $s$ for different values of $\beta$. $s$ is plotted against two types of the theoretical learning speed $s_\text{th}$ of  \eqref{eq:lrnspd_spneig} and $s'_\text{th}$  of  \eqref{eq:lrnspd_spnvar} in cross and circle markers, respectively.
    Different color codes data for different $\beta$ as shown on the color bar. The same color code is in Panels B and C.
    B: $s$ for 5 maps that have the same input are shown against the variance of the spontaneous fluctuation along the target directions in (i), while $s$ for 5 maps that have the same target are plotted against the response and the spontaneous fluctuation along the input direction in (ii) and (iii), respectively.
    These data are obtained for $\beta=0.6$.
    C: $T^{-1}_L$ is plotted against $s_\text{th}$ by cross and $s'_\text{th}$ by circle.
    }
    \label{fig:lrnspd_rnd_rndJ}
\end{figure*}

In the main text, we have analyzed the input and target patterns parallel to the eigenvectors in the random network model (eigenvector maps).
Here, we verify the formulae  \eqref{eq:lrnspd_spneig} 
 and \eqref{eq:lrnspd_spnvar}  in the case of the random input and target patterns (random maps). 
In the random maps, the derivation of the learning speed  $s_\text{th}$ is not exact compared with the case for the eigenvector maps, and the validity of these formulae is not completely assured.
Thus, we verify these formulae numerically.
We set input and target patterns to be random as given by $\pm$  binary vectors whose elements are generated according to the probability distributions $P(\xi_i = \pm 1)=P(\eta_i = \pm 1)=0.5$.

To verify the formula  \eqref{eq:lrnspd_spneig}, we computed the learning speed $s$ and the theoretical learning speed $s_\text{th}$ determined by this formula.
Figure \ref{fig:lrnspd_rnd_rndJ}A demonstrates that $s$ agrees well with $s_\text{th}$ for various values of $\beta$. 
Theoretical speed $s_\text{th}$ in Eq. \eqref{eq:lrnspd_spneig} is dependent on the variance of the spontaneous dynamics $Var_{\boldsymbol{\xi}}( \boldsymbol{x})$ and the response $|\boldsymbol{x_r}|$.
To examine explicitly how each of the two factors contributes to the learning speed, the measured speed $s$ is plotted as functions of  $Var_{\boldsymbol{\xi}}( \boldsymbol{x})$  and  $|\boldsymbol{x_r}|$ separately in Fig. \ref{fig:lrnspd_rnd_rndJ}B.
$s$  is linearly proportional to  $Var_{\boldsymbol{\xi}}( \boldsymbol{x})$  and  $|\boldsymbol{x_r}|^2$  well, which agrees with the theoretically obtained formula between them in \eqref{eq:lrnspd_spneig}.

Next, to examine if the formula  \eqref{eq:lrnspd_spnvar} on $s_{\text{th}}'$ is valid,
we compare directly $s$ with $s'_\text{th}$ in Fig. \ref{fig:lrnspd_rnd_rndJ}A, which confirms this formula.
In the same manner as Fig. \ref{fig:lrnspd_rnd_rndJ}B(i)(ii), we, further, computed the contribution of $Var_{\boldsymbol{\eta}}( \boldsymbol{x})$ to the learning speed and found that the learning speed is proportional to $Var_{\boldsymbol{\eta}}( \boldsymbol{x})$ as shown in Fig. \ref{fig:lrnspd_rnd_rndJ}B(iii).

Finally,  we confirm that $T^{-1}_L$ is inversely proportional to $s_\text{th}$ and $s'_\text{th}$ in the random maps in Fig \ref{fig:lrnspd_rnd_rndJ}C.

\section*{Appendix 3: the random maps in the pre-embedded network model}

\begin{figure*}[t]
    \centering
    \includegraphics[width=0.7\linewidth]{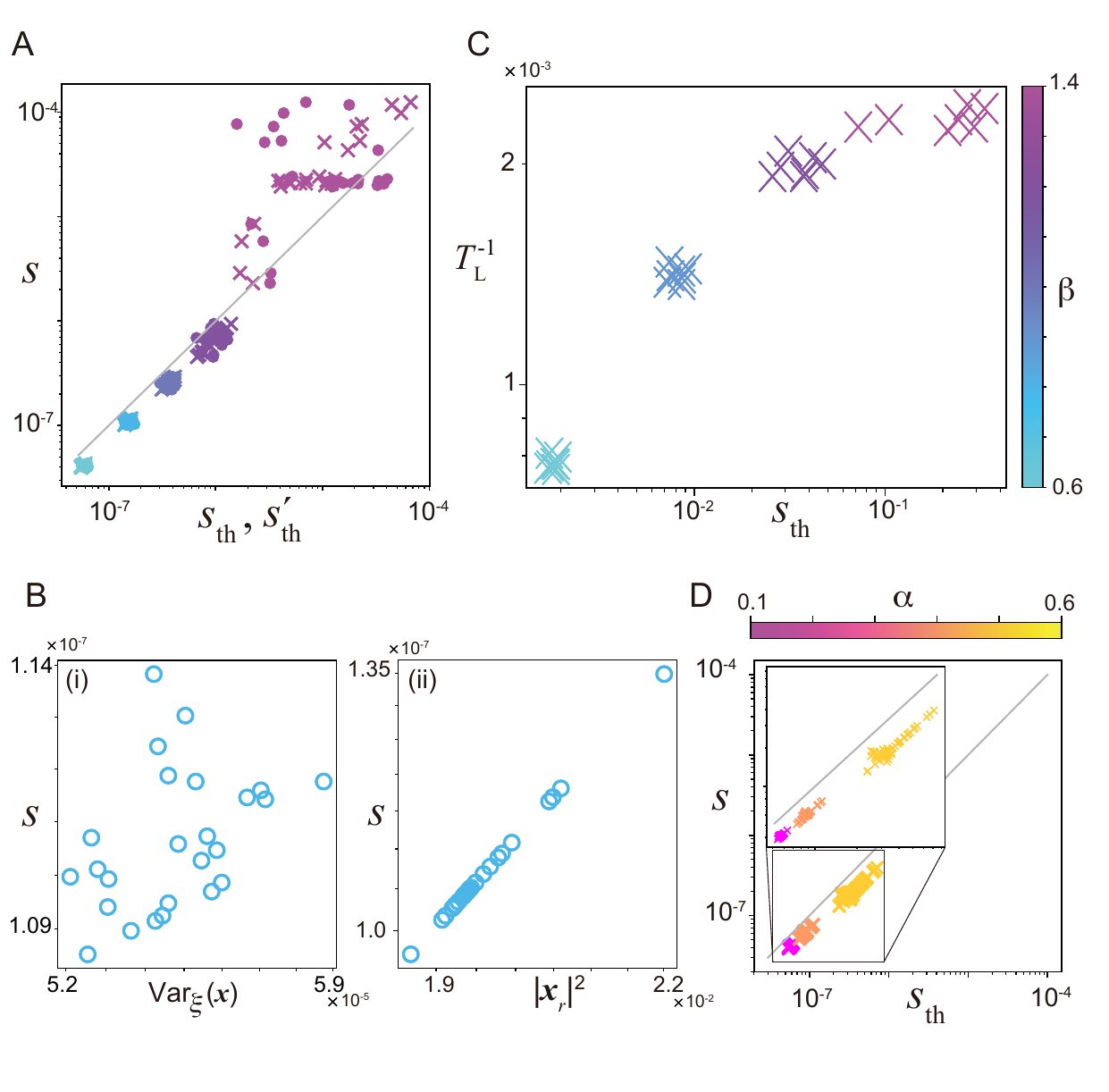}
    \caption{The learning speed in the random maps in the pre-embedded network model. 
    A: The measured learning speed $s$ for different values of $\beta$ is plotted against $s_{\text{th}}$ and  $s'_{\text{th}}$ in the same manner as in Fig. \ref{fig:lrnspd_randJ}A. Different colors represent the learning speed for different values of $\beta$, as indicated by the color bar. 
    B: For $\beta=0.8$, $s$ is shown against the variance of the spontaneous fluctuation along the target direction in (i) and against the response in (ii), respectively. 
       C: $T^{-1}_L$ is plotted against $s_\text{th}$.
       D: $s$ for different values of $\alpha$ is plotted against $s_\text{th}$.}
    \label{fig:lrnspd_rnd_embJ}
\end{figure*}

In this Appendix, we examine the fluctuation-learning relationship in the random maps in the pre-embedded network model.
First, we verify the formula \eqref{eq:lrnspd_spneig} by comparing the learning speed $s$ with the theoretical one $s_\text{th}$ for $\alpha=0.1$ as in Fig. \ref{fig:lrnspd_rnd_embJ}A and find that $s$ agrees with $s_\text{th}$ for almost all of $\beta$ except that for $\beta=1.4$, which is close to the critical value of $\beta$ exhibited in Fig. \ref{fig:spn_all}A.

Next, the contributions of the two factors, $Var_{\boldsymbol{\xi}}( \boldsymbol{x})$ and $\boldsymbol{x_r}$, to the learning speed $s$, are explored separately.
$Var_{\boldsymbol{\xi}}(\boldsymbol{x})$ was computed for $25$ maps that have the identical input, whereas  $\boldsymbol{x_r}$ was for $25$ maps that have the identical target in Fig. \ref{fig:lrnspd_rnd_embJ}B.
$s$ is well proportional to $|\boldsymbol{x_r}|^2$  and the proportionality to $Var_{\boldsymbol{\xi}}(\boldsymbol{x})$ is not so good.
This weak proportionality might be attributed to much smaller fluctuation along the random pattern direction than that along the target or input pattern direction (Fig. \ref{fig:spn_all}) that leads to a narrower distribution of $s_\text{th}$.

Further, the learning time to complete $T_L$  is explored.
We plot  $T^{-1}_L$ against $s_\text{th}$  in Fig. \ref{fig:lrnspd_rnd_embJ}C.
As shown, $T^{-1}_L$ is inversely proportional to $s_\text{th}$.
Thus, the learning time to complete is also estimated well by the spontaneous fluctuation in the direction of the target and the response in the pre-embedded network model.

Finally,  the dependence of the learning speed on $\alpha$ is investigated.
Fig. \ref{fig:lrnspd_rnd_embJ}D exhibits that the learning speed $s$ increases with  $\alpha$ in agreement with the theoretical value $s_\text{th}$.
This result shows that the formula \eqref{eq:lrnspd_spneig} is valid for different $\alpha$ even in the random map case.

\section*{Appendix 4: fluctuation-learning relationship in Hopfield network model with Hebb rule}

To validate the fluctuation-learning relation for auto-associative memory, we analyze Hopfield network model, $J=\Sigma_\mu \boldsymbol{\xi}^{\mu}(\boldsymbol{\xi}^{\mu})^T/N, (\mu = 1,\cdots, \alpha N)$, where $\boldsymbol{\xi}$ is a random binary $N$-dimensional vector as a target.
The network learns a new pattern $\boldsymbol{\xi}^{\mu+1}$ with Hebb learning rule 
\begin{align}
    \tau_J \Delta J = (\boldsymbol{x}\boldsymbol{x}^T/N)\Delta t,\label{eq:lrn_hebb}
\end{align}
in the presence of the input $\boldsymbol{\xi}^{\text{new}}$.
Here, the target $\boldsymbol{\xi}^{\text{new}}$ is applied as an input pattern in the auto-associative memory.
The applied input needs to be sufficiently strong for storing a new pattern, 
otherwise, the Hebb learning rule cannot destabilize the present state to memorize a new pattern.
We set $\gamma=0.5$, which is much stronger than that in the simulations of the I/O mapping model.

\begin{figure*}
    \centering
    \includegraphics[width=0.7\linewidth]{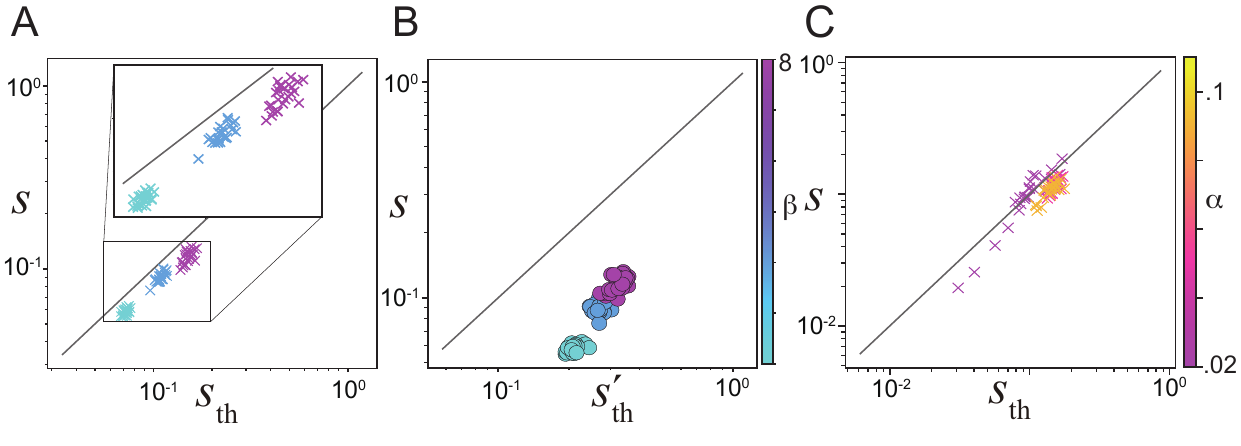}
    \caption{The spontaneous fluctuations and the learning speed in the Hopfield model.
    A and B: The measured learning speed $s$ for different values of $\beta$ is plotted against $s_{\text{th}}$ and $s'_{\text{th}}$ in A and B, respectively, in the same manner as in Fig. \ref{fig:lrnspd_randJ}A. Different colors represent the learning speed for different values of $\beta$ as shown in the color bar.
    $\alpha$ is set at $0.05$.
    C: The measured learning speed $s$ for different values of $\alpha$ is plotted against $s_{\text{th}}$ for $\beta=6$.
    }
    \label{fig:lrnspd_hopf}
\end{figure*}

In the Hopfield network model, the embedded targets are stable fixed points when the number of memories is below the critical memory capacity for infinitely large $\beta$ (the fixed points are not exactly the target for a finite value of $\beta$, but they are close to the targets).
Therefore, we analyze the neural dynamics and the learning speed after convergence into one of these stable fixed points, denoted as $\boldsymbol{\xi}^0$.
In contrast to the pre-embedded model, only the random case is considered and the remap case is not considered in the Hopfield model due to its setup mentioned above.

In this model with $N$ neurons, the neural dynamics (Eq. \eqref{eq:neuro_dyn1}) are the same as those studied in the main text, whereas the learning rule and the connectivity matrix $J$ are different from those in the main text.
We obtain the learning speed in Hopfield network model as
\begin{align}
    \frac{\Delta \boldsymbol{x}^*}{\Delta t} =(\frac{|\boldsymbol{x_r}|^2}{DN\tau_J})  Cov(\boldsymbol{x})_{\text{inp}}B \boldsymbol{x}_r,
\end{align}
by replacing the perceptron-like learning rule in Eq. \eqref{eq:lrn_rule} with the Hebb rule in Eq.\eqref{eq:lrn_hebb} and by substituting $\boldsymbol{x_r}$ to $\boldsymbol{x}$ for the initial state of learning.
By approximating the covariance matrix of the fluctuations in the presence of the input $Cov(\boldsymbol{x})_{\text{inp}}$ by the spontaneous fluctuation around $\boldsymbol{\xi}^0$,  $Cov(\boldsymbol{x})$,
the formula corresponding to  \eqref{eq:lrnspd_res} is obtained as
\begin{align}
     \frac{\Delta \boldsymbol{x}^*}{\Delta t} =(\frac{|\boldsymbol{x_r}|^2}{DN\tau_J})  Cov(\boldsymbol{x})B \boldsymbol{x}_r.
\end{align}
After approximating  $Cov(\boldsymbol{x})$ by $Var_{B\boldsymbol{x_r}}(  \boldsymbol{x}) $ in the same manner as I/O mapping models,  we derive
\begin{align}
    \frac{\Delta \boldsymbol{x}^*}{\Delta t} =(\frac{|\boldsymbol{x_r}|^2}{DN\tau_J})  Var_{B\boldsymbol{x_r}} 
 (\boldsymbol{x}) B \boldsymbol{x_r}.\tag{FLR2''} \label{eq:lrnspd_hebb}
\end{align}
This means that the learning speed of associative memory is represented by the variance of the spontaneous fluctuations and the response in the same manner as I/O mapping models.

As the input is large and $\boldsymbol{x_r}$ is far from the spontaneous state in the Hopfield network model, the validity of a formula corresponding to  \eqref{eq:lrnspd_spnspn} would be questionable. 
Still, one could derive it by linearizing the neural dynamics in Eq. \eqref{eq:neuro_dyn1} around $J\boldsymbol{\xi}^0$, instead of the origin, we obtain
\begin{align}
    \frac{d (\boldsymbol{x}-\boldsymbol{\xi}^0)}{dt} = \phi(J\boldsymbol{\xi}^0) + \phi'(J\boldsymbol{\xi}^0) (J(\boldsymbol{x}-\boldsymbol{\xi}^0)+\gamma \boldsymbol{\eta})-\boldsymbol{x} + \boldsymbol{\zeta}.
\end{align}
By using $\boldsymbol{\xi}^0 = \phi(J\boldsymbol{\xi}^0)$, we have the fixed point of this linearized dynamics after averaging over time that corresponds to the response to $\boldsymbol{\eta}$,  $\boldsymbol{x_r}$, as 
\begin{align}
    \boldsymbol{x_r} - \boldsymbol{\xi}^0 &=& \gamma (1-\phi'(J\boldsymbol{\xi}^0)J)^{-1} \phi'(J\boldsymbol{\xi}^0) \boldsymbol{\eta}\\
    &=& \frac{\gamma}{D} Cov(\boldsymbol{x}) B' \boldsymbol{\eta},
\end{align}
where $\phi'(J\boldsymbol{\xi}^0)$ is denoted by $B'$.
In the same manner as in the subsection "Learning speed for Hebb-type rule",  we obtain the representation of the response as 
\begin{align}
    \boldsymbol{x_r} = \boldsymbol{\xi}^0 + \frac{\gamma}{D} Var_{B' \boldsymbol{\eta}} (\boldsymbol{x}) B' \boldsymbol{\eta}. \label{eq:resp_hopf}
\end{align}
Finally, the formula corresponding to \eqref{eq:lrnspd_spnspn} is obtained by substituting Eq.  \eqref{eq:resp_hopf} to  \eqref{eq:lrnspd_hebb},
\begin{multline}
     \frac{\Delta \boldsymbol{x}^*}{\Delta t} =(\frac{|(\boldsymbol{\xi}^0 + \frac{\gamma}{D} Var_{B' \boldsymbol{\eta}} (\boldsymbol{x}) B' \boldsymbol{\eta})|^2}{DN\tau_J}) \\
     \times Var_{B' \boldsymbol{\eta}} (\boldsymbol{x}) B (\boldsymbol{\xi}^0 + \frac{\gamma}{D} Var_{B' \boldsymbol{\eta}} (\boldsymbol{x}) B' \boldsymbol{\eta}).\tag{FLR3''} \label{eq:lrnspd_hebb_spn}
\end{multline}
\eqref{eq:lrnspd_hebb} and \eqref{eq:lrnspd_hebb_spn} show that the learning speed of the associative memory is also represented in the same manner as I/O mapping models.

These equations are obtained by rough approximations. 
We numerically verify these relationships \eqref{eq:lrnspd_hebb} and \eqref{eq:lrnspd_hebb_spn}.
First, we numerically confirmed  \eqref{eq:lrnspd_hebb} for various values of $\beta$.
Figure \ref{fig:lrnspd_hopf}A demonstrates that the measured learning speed $s$ agrees rather well with the theoretical learning speed $s_\text{th}$ determined by  \eqref{eq:lrnspd_hebb}.
As $\beta$ increases, the learning speed increases, as in the random network model and the pre-embedded model.

Next, we explored whether the measured learning speed agrees with the theoretical speed $s_{\text{th}'}$ determined by \eqref{eq:lrnspd_hebb_spn}, as shown in Fig. \ref{fig:lrnspd_hopf}B.
The two types of speed,  $s$ and $s_{\text{th}'}$, are correlated, but the difference between them is larger than those in other models.
As already mentioned, the applied input in the Hopfield network model is much larger than I/O mapping models, so the assumption for \eqref{eq:lrnspd_hebb_spn} is not satisfied.
This leads to the large deviation between the measured and theoretical learning speeds, but they are still proportional to each other.

Finally, the dependence of the learning speed on the number of embedded memories $\alpha$ is investigated.
We measured the learning speed $s$ and computed the theoretical one $s_{\text{th}}$ for the various values of $\alpha$ and plotted them in Fig. \ref{fig:lrnspd_hopf}C.
The numerical result agrees with the formula \eqref{eq:lrnspd_hebb} for all values of $\alpha$.
For lower $\alpha$, here $0.02$, the variety of $s$ across the different targets to be learned is broader.
The applied input $\boldsymbol{\eta}$ that is identical to the target evokes the response of the network dynamics through the interaction between $J \boldsymbol{\eta} = \sum_\mu (\boldsymbol{\xi}^{\mu}) (\boldsymbol{\xi}^{\mu})^T \boldsymbol{\eta}/N$.
Thus, for low $\alpha$, the interactions across different inputs vary more due to the small number of the overlaps $(\boldsymbol{\xi}^{\mu})^T \boldsymbol{\eta}/N$, leading to the large variety of the responses and, consequently, the large variety of the learning speed.

To sum up, our results show that the formula of the fluctuation-learning relation in \eqref{eq:lrnspd_hebb} is verified for various values of $\beta$ and $\alpha$ in the Hopfield model.

\end{document}